\documentclass[10pt]{article}
\usepackage{latexsym}
\usepackage{amssymb}
\usepackage{amsfonts}
\usepackage{amsmath}
\usepackage{theorem}

\theorembodyfont{\upshape}

\newtheorem{theorem}{Theorem}

\newtheorem{definition}{Definition}

\begin{document}

\title{Structural Properties of the Disordered\\ 
Spherical and other Mean Field Spin Models}
\author{Luca De Sanctis
\footnote{ICTP, Strada Costiera 11, 34014 Trieste, Italy,
{\tt<lde\_sanc@ictp.it>}}}

\maketitle

\begin{abstract}
We extend the approach of Aizenman, Sims and Starr for the SK-type
models to their spherical versions. Such an extension has already
been performed for diluted spin glasses. The factorization property
of the optimal structures found by Guerra for the SK model, which holds
for diluted models as well, is verified also in the case of spherical systems,
with the due modifications.
Hence we show that there are some common structural features in 
various mean field spin models. These similarities seem to 
be quite paradigmatic, and we summarize the various techniques
typically used to prove the structural analogies and to 
tackle the computation of the free energy per spin in the 
thermodynamic limit.
\end{abstract}



\section{Introdution}

Aizenman, Sims and Starr introduced in \cite{ass} an approach 
to models of the type of Sherrington-Kirkpatrick (SK) 
of mean field spin glasses which is based on very
physical ideas. Their method consists of introducing 
an auxiliary system to be coupled to the
original one, and use it to generate trial order parameters and to find a
general form for the trial free energy that
bounds the actual one. They also find that the existence of thermodynamic
limit of the free energy suggests the expression of an optimal 
trial free energy, which is difficult to compute but ensures an Extended Variational
Principle: the exact free energy is obtained as an infimum in a suitable space.
This approach explains in a sense the structure of the model: its main parameters,
the form of the trial free energy, the trial order parameters, the probability
space in which one has to search for the optimal trial free energy. In other words,
the method tells us what is the correct physical approach to the model, i.e.
coupling it to a general auxiliary system with the proper features.
Guerra found in \cite{g1} a way to restrict the space in which one has
to work by proving some invariance properties of the optimal structures.
The whole approach seemed to rely heavily on the Gaussian nature
of the couplings. On the contrary, being a very physical approach, 
it turned out to be quite paradigmatic, once the proper techniques 
for various other models were found (\cite{lds1, lds3}). 
As a matter of fact, not only the same approach works when the model
is diluted and loses the Gaussian nature of the couplings, but we prove
here that making the model spherical either does not affect its structure.
The most important starting point has to be a simpler one, though. If the 
method is paradigmatic it must work first of all in the simple case
of the non-disordered version of the SK model: the mean field 
ferromagnet, that should be analogous to
disordered systems from the structural point of view.
This is quite the case (\cite{g2}), of course, as we will see.

The main idea is always to replace the quadratic dependance of the
Hamiltonians on the main physical quantity (magnetization, overlap, etc.)
by a linear approximation with suitable coefficients, generated by
the auxiliary structure. Hence one compares quadratically the 
main physical quantity of the original system with the trial one coming
from the auxiliary system, and this leads to a free energy given by
the difference of an entropy term and an internal energy term.
This offers important results for a large class of models, 
but one needs to find the proper
technique to prove analogous theorems for different models. 

We will start with the description of the approach in the case of the 
Curie-Weiss (CW) mean field ferromagnet, which provides a
stereotypical example. We then show how the same method applies
to the SK model and its spherical version.
Lastly, we explain how and why the same results extend to diluted models,
including Optimization problems.

%

\section{Mean Field Model of Ferromagnets}

The Curie-Weiss Hamiltonian is
\begin{equation*}
H_N^{CW}(\sigma)=-\frac JN\sum_{i<j}^{1,N}\sigma_i
\sigma_j\ ,
\end{equation*}
defined on Ising spin configurations 
$\sigma:i\to\sigma_i=\pm1\ , i=1,\ldots,N$. 
The coupling strength $J$ is a positive constant.
Defining the magnetization per 
spin of the configuration $\sigma$ as
\begin{equation*}
\label{magnetizzazione}
m(\sigma)=\frac1N\sum_{i=1}^N\sigma_i\ ,
\end{equation*}
the Hamiltonian can then be written as
\begin{equation*}
H_N^{CW}(\sigma)= - N\frac12J m^2(\sigma)\ ,
\end{equation*}
so that the partition function is by definition
\begin{equation*}
Z_N(\beta)=\sum_{\{\sigma\}}\exp (N\frac{1}{2}\beta Jm^2(\sigma))\ ,
\end{equation*}
and the free energy is
\begin{equation*}
-\beta f = \lim_N -\beta f_N =\lim_N \frac1N\ln Z_N\ ,
\end{equation*}
omitting the dependence on $\beta$.
Notice the following fundamental formula and keep it in mind for later
\begin{equation}
\label{fundamental1}
\partial_\beta \frac1N \ln Z_N(\beta)=\frac12 J\langle m^2 \rangle\ ;
\end{equation}
here $\langle \cdot \rangle$ denotes the Boltmann-Gibbs average.

\subsection{Magnetization Structures and the Comparison Method}

Consider some discrete space $\Sigma$, some weights 
$\xi_\tau, \tau\in \Sigma$ and a magnetization kernel 
$\tilde{m}:\Sigma^2\to [-1,1]$. The triple $(\Sigma, \xi, \tilde{m})$
is called {\em Magnetization Structure} (MaSt). 
Then for a given MaSt $\mathcal{M}$ define the trial function
\begin{equation}
\label{trial}
G_N(\mathcal{M})=
\frac1N \ln \frac{\sum_{\sigma,\tau}\xi_\tau\exp(\beta JNm\tilde{m})}
{\sum_\tau\xi_\tau\exp(\frac12\beta JN \tilde{m}^2)}
\end{equation}
The idea is to compare the trial magnetization $\tilde{m}$
with the actual magnetization of the original system.
In the simple case of the CW model we do not need such a rich structure,
but it serves as an introduction to the methods. For a very formal
treatment of such methods for classical spin systems, an excellent
reference is \cite{ks}.


\subsubsection{Generalized MaSt bound}

Consider the following special MaSt $\mathcal{M}_{\tilde{m}}$. 
Let $\Sigma$ be any space
and $\xi$ be any weights in this space.
Take an auxiliary CW system of $M$ spins, with one-body interactions, 
i.e. with
Hamiltonian linear in the magnetization. Assume the coupling is still $J$,
and the temperature is $\tilde{\beta}$. It is easy to compute the mean
magnetization $\tilde{m}$ in this system
\begin{equation*}
\tilde{m}= \tanh{\tilde{\beta} J}\ ,
\end{equation*}
which we can modulate between zero and one by changing the temperature
(or equivalently the strength of the couplings). 
Now take this fixed value $\tilde{m}$
(constant over all $\Sigma$) as the trial 
magnetization of our MaSt $\mathcal{M}_{\tilde{m}}$, 
and plug it into (\ref{trial}). We get
\begin{equation}\label{g}
G_N(\mathcal{M}_{\tilde{m}})=
\frac1N\ln\sum_\sigma\exp(\beta J N m\tilde{m})-
\frac1N\ln\exp(\frac12\beta J N \tilde{m}^2)
\end{equation}
since $\tilde{m}$ does not depend on the points $\tau\in\Sigma$.
Notice that $NG_N(\mathcal{M}_{\tilde{m}})$ 
coincides with the logarithm of the partition function 
$Z(H_N^{CW}(m^2))$,
provided we replace the squared magnetization $m^2$ with the trivial 
quadratic estimate
$m^2 \geq 2m\tilde{m}-\tilde{m}^2$
which, as $G_N(\mathcal{M}_{\tilde{m}})$ is easy to compute, 
brings the consequent inequality, holding for all $N$
\begin{equation}\label{cwbound1}
\frac1N \ln Z_N(\beta)\geq \sup_{\tilde{m}}
\{\ln 2+\ln\cosh(\beta J \tilde{m}) - \frac12 \beta J \tilde{m}^2\}
=\sup_{\tilde{m}}G_N(\mathcal{M}_{\tilde{m}})\ .
\end{equation}
Notice that $G_N(\mathcal{M}_{\tilde{m}})$ does not depend on 
$N$.
Physically speaking, we replaced the 
two-body interaction, which is difficult to deal with, with a one-body interaction.
Then we try to compensate this by modulating the field acting on each spin
by means of a trial fixed magnetization (entropy term), and a 
correction term (the internal energy), quadratic in this
trial magnetization $\tilde{m}$.  

In general, we can state (see also \cite{ks} for a 
different perspective) the following
\begin{theorem}
For any 
value of $N\in\mathbb{N}$, the trial function 
$G_N$ defined by (\ref{trial}) is a lower bound
for the pressure $-\beta f_N$ of the CW model
\begin{equation}\label{b1}
-\beta f_N \geq \sup_{\mathcal{M}}G_N(\mathcal{M}) \ ,
\end{equation}
\end{theorem}
Let us emphasize that a fundamental property of the CW model is that,
for any $\tilde{m}$ (non constant on $\Sigma$ in general),
the inequality 
\begin{equation}
\label{main1}
m \geq 2m\tilde{m}-\tilde{m}^2
\end{equation}
implies a correspondent inequality for the free energy per spin.

Another fundamental application of the fact that convex inequalities 
for the magnetization imply inequalities for the free energy is the
proof of the existence of the thermodynamic limit of the free energy per spin
(\cite{g3, ks}).
The proof relies on considering two CW systems, one with $N$ spins
and another one with $M$ spins, then comparing these two independent
systems with the system consisting of the union of the two, which is a CW model
with $N+M$ spins.

\subsubsection{Reversed bound and variational principle}

The just mentioned existence of the thermodynamic limit of the free
energy density, guarantees that the following Ces{\`a}ro limit offers
the value of such a limit
\begin{equation}\label{boltz}
\mathbf{C}\lim_M\frac1N\ln\frac{Z_{M+N}}{Z_M}= -\beta f\ .
\end{equation}
Now assume $M>>N$. Clearly $Z_{M+N}$ splits into 
the sum of three terms. One contains the 
interactions between spins in the big system, one contains the 
interactions between one spin in the big system and one in the small system,
one contains the interactions between spins in the small system.
The latter is negligible (by convexity arguments). 
This means that the spins in the small
system are decoupled by the addition of the large system, and they do not
interact one another (in the $M$-limit).
If we take the term with the interactions within the large system (or cavity)
as the weight $\xi_\tau$, $\tau$ as the spin
configuration in $\Sigma=\{-1, +1\}^M$, then the left hand side of (\ref{boltz})
is the trial function $G_N(\mathcal{M}_B(M))$ (which does not
depend on $N$) with the special choice
just described, called Boltzmann MaSt $\mathcal{M}_B(M)$. 
In fact, the Boltzmann factor
in the denominator with $Z_N$ can be written as $\xi$ times a properly
renormalized CW Hamiltonian, like in (\ref{g}). So we get the following
\begin{theorem}[Reversed Bound]
If we plug the Boltzmann MaSt $\mathcal{M}_{B}(M)$ 
just defined above into the 
trial function $G_N$ defined by (\ref{trial}),
then the Boltzmann trial function $G_{N}(\mathcal{M}_{B}(M))$
provides, in the thermodynamic limit, an upper 
bound for the CW pressure $-\beta f$
\begin{equation*}
-\beta f\leq\lim_{N\rightarrow\infty}\limsup_{M\rightarrow\infty}
G_{N}(\mathcal{M}_{B}(M))=\lim_{N\rightarrow\infty}G_{N}(\mathcal{M}_{B})\ ,
\end{equation*}
\end{theorem}
as a consequence of
\begin{equation}\label{rb}
-\beta f=\mathbf{C}\lim_M\frac1N\ln\frac{Z_{M+N}}{Z_M}\leq
\limsup_N\limsup_M G(\mathcal{M}_B(M))\equiv G(\mathcal{M}_B)\ .
\end{equation}
Therefore we have the reversed bound to (\ref{b1}), and thus
the following
\begin{theorem}[Extended Variational Principle] Taking the supremum,
for each $N$ separately, of the trial function 
$G_N(\mathcal{M})$ over the whole
MaSt space, the resulting sequence tends to the limiting pressure 
$-\beta f$ of the CW model as $N$ tends to infinity
\begin{equation}
\label{evp-f}
-\beta f = \lim_N\sup_{\mathcal{M}}G_N(\mathcal{M})\ .
\end{equation}
\end{theorem}

In order to compute explicitly $f$, one can proceed as follows (\cite{g2}).
Let us notice that the magnetization $m$ can take only $2N+1$ distinct values.
We can therefore split the partition function into sums over configurations with
constant magnetization in the following way
\begin{equation}\label{split}
Z_N(\beta)=\sum_\sigma\sum_{\tilde{m}}
\delta_{m\tilde{m}}e^{\frac12 \beta J m \tilde{m} N}\ \ \Leftarrow\
\sum_{\tilde{m}}\delta_{m\tilde{m}}=1\ .
\end{equation}
Now inside the sum $m=\tilde{m}$, or better $m^2=\tilde{m}^2$, 
which means also $m^2=2m\tilde{m}-\tilde{m}^2$.
Plugging such a quadratic {\sl equality}
into $Z_N(\beta)$ and using the trivial inequality 
$\delta_{m\tilde{m}}\leq 1$
yields
\begin{equation*}
Z_N(\beta)\leq\sum_{\tilde{m}}\sum_\sigma e^{\beta J Nm\tilde{m}}
e^{-\frac12 \beta J N \tilde{m}^2}\ .
\end{equation*}
But this clearly means that
\begin{equation*}
Z_N(\beta)\leq\sum_{\tilde{m}}\sup_{\tilde{m}}
\{\ln 2+\ln\cosh(\beta J \tilde{m}) - \frac12 \beta J \tilde{m}^2\}
\end{equation*}
from which
\begin{equation*}
\frac1N \ln Z_N(\beta)\leq\frac{\ln (2N+1)}{N}+\sup_{\tilde{m}}
\{\ln 2+\ln\cosh(\beta J \tilde{m}) - \frac12 \beta J \tilde{m}^2\}\ .
\end{equation*}
This gives, together with (\ref{cwbound1}),
the exact value of free energy per site at least in the thermodynamic limit.
The presence of the correction term $\ln (2N+1)/N$ is typical of these kinds of bounds,
and as a consequence one can usually get the exact value of the free energy
only in the thermodynamic limit (but compare to Proposition 5 in \cite{ks}
for the special case of classical spin systems).

\subsection{Spherical Ferromagnetic Model}

The calculation of the free energy of the 
spherical version of the CW model is more involved than the Ising case
just illustrated. Still, the computation can be performed (\cite{bk}) 
and the it can be reproduced in the framework of the Magnetization Structures,
but we will present here in detail only the disordered case.

\section{Disordered Mean Field Models of Spin Glasses}

The Hamiltonian of the Sherrington-Kirkpatrick (SK) model 
comes from the CW one where we take independent centered unit
Gaussian couplings $\{J_{ij}\}$ as opposed to constant ones,
after a proper rescaling:
\begin{equation}
\label{SK}
H_N(\sigma,h;J)=-\frac1{\sqrt N} \sum_{i<j}^{1,N} J_{ij} \sigma_i\sigma_j\ .
\end{equation}
The Hamiltonian is in some sense still quadratic in the 
quantity that takes the place of the magnetization: being a centered
Gaussian, the Hamiltonian is determined by its covariance
\begin{equation*}
\mathbb{E}(H(\sigma)H(\sigma^{\prime}))=Nq^2(\sigma, \sigma^{\prime})\ ,\
q(\sigma, \sigma^{\prime})=\frac1N\sum_{i=1}^N\sigma_i\sigma^{\prime}_i
\end{equation*}
where $\mathbb{E}$ will denotes the expectation with respect to the $J_.$'s,
and $q$ is the overlap between two {\sl replicas}.
Notice (and compare with the analog in the CW model) that
\begin{equation}
\label{fundamental2}
\partial_\beta \frac1N \mathbb{E}\ln Z_N(\beta)=
\frac{\beta}{2}(1-\langle q^2 \rangle)\ ,
\end{equation}
where $\langle \cdot \rangle$ denotes the 
composition of the Boltmann-Gibbs ($\Omega$) average (taken first),
followed by the {\sl quenched} expectation ($\mathbb{E}$).
The key tool employed to make the calculation above is the
{\sl Gaussian integration by parts}.

As the SK model is random (because of the random couplings) and
the overlap plays a role similar to that of the magnetization for the CW model,
it is expected that the analog of the MaSt for disordered model is the 
Random Overlap Structure (ROSt), given in the following
\begin{definition}
A \emph{Random Overlap Structure} 
$\mathcal{R}$ is a triple 
$(\Sigma, \tilde{q}, \xi)$ where  
\begin{itemize}
  \item $\Sigma$ is a discrete space;
  \item $\xi: \Sigma\rightarrow\mathbb{R}_+$ 
  is a system of random weights; 
  \item $\tilde{q}:\Sigma^2\rightarrow[0, 1] , |\tilde{q}|\leq 1$
  is a  positive definite \emph{Overlap Kernel} 
  (equal to 1 only on the diagonal of
  $\Sigma^2$).
\end{itemize}
\end{definition}

Now consider two families of independent centered Gaussian random variables 
$\tilde{H}_{.}$ and $\hat{H}$, defined on $\Sigma\ni \gamma$, such that
\begin{equation}\label{covarianze}
\mathbb{E}(\tilde{H}_{i}(\gamma)\tilde{H}_{j}(\gamma^{\prime}))
=N2\tilde{q}_{\gamma\gamma^{\prime}}\delta_{ij}\ ,\ 
\mathbb{E}(\hat{H}(\gamma)\hat{H}(\gamma^{\prime}))=
N\tilde{q}^{2}_{\gamma\gamma^{\prime}}
\end{equation}
and consider 
$\tilde{H}=\sum_{i=1}^{N}\tilde{H}_{i}\sigma_{i}$. 
Then we define, in analogy with (\ref{trial}), the \emph{Generalized Trial Function} by
\begin{equation}\label{gtf}
G_{N}(\mathcal{R}, \tilde{H}, \hat{H})=\frac 1N\mathbb{E}\ln
\frac{\sum_{\sigma, \tau}\xi_{\tau}
\exp(-\beta \tilde{H})}{\sum_{\tau}\xi_{\tau}
\exp(-\beta \hat{H})}\ .
\end{equation}
At this point we can use interpolation (\cite{g3})
to prove (\cite{ass}) the following
\begin{theorem}[Generalized Bound]
\label{b2}
For any 
value of $N\in\mathbb{N}$, the trial function $G_N$ defined by (\ref{gtf}) is an upper bound
for the pressure $-\beta f_N$ of the SK model
\begin{equation*}
\label{ }
-\beta f_N\leq 
\inf_{\mathcal{R}} G_N\ .
\end{equation*}
\end{theorem}
The proof can be found in \cite{ass}.
To see why this inequality has the same nature as the analogous one
for the CW model, i.e. 
(\ref{b1}), we have to recall (\ref{fundamental2}) and notice that,
thanks to a differentiation with respect to the interpolating parameter,
the theorem above is guaranteed by the analog of (\ref{main1})
for the covariances of the three Hamiltonians $H,\tilde{H},\hat{H}$ 
we used
\begin{equation}
\label{main2}
q^2 \leq 2q\tilde{q} - \tilde{q}^2\ .
\end{equation}
The same result for the CW model could be proven by interpolation too,
but that case is so simple that interpolation would have been a further unnecessary complication.
In order to emphasize even more the key role of (\ref{main2}),
like for the CW, let us understand that the theorem above is 
part of the following general comparison scheme (see \cite{g2, g3} and
references therein).
\begin{theorem}
Let $U_i$  and $\hat{U}_i$, for $i=1,\ldots,K$, be independent families of
centered Gaussian random variables, whose covariances satisfy
the inequalities for generic configurations
$$
\mathbb{E}(U_iU_j)\equiv S_{ij}\geq
\mathbb{E}(\hat{U}_i\hat{U}_j)\equiv \hat{S}_{ij}\ ,
\mathbb{E}(U_iU_i)\equiv S_{ii}=
\mathbb{E}(\hat{U}_i\hat{U}_i)\equiv \hat{S}_{ii}\ ,
$$
then for the quenched averages we have the inequality in the opposite
sense
$$
\mathbb{E}\ln\sum_iw_i\exp(U_i)\leq \mathbb{E}\ln\sum_iw_i\exp(\hat{U}_i)
$$
where the $w_i\geq 0$ are the same in the two expressions.
\end{theorem}
The proof (\cite{g2, g3}) is very simple, and based on interpolation.
A very important application, other than the ROSt approach - which
includes as a special realization the Parisi trial function, is the 
proof of the existence of thermodynamic limit of 
the free energy density (\cite{gt3, g3}).

Now take $\Sigma=\{-1,1\}^M$, and denote by $\tau_.$ the elements of $\Sigma$.
We clearly have in mind an auxiliary spin systems. In fact,
we also choose
\begin{equation*}
\tilde{H}_.=-\sum_{k=1}^M J^k_.\tau_k\ ,\ \ \ 
\hat{H}=-\sqrt{\frac{N}{2}}\sum_{k,l}^{1,M}J^{k,l}\tau_k\tau_l
\end{equation*}
which satisfy (\ref{covarianze}). Let us also chose
$\xi_\tau=\exp(-\beta H^{SK}_M(\tau))$.
Then, if we call $\mathcal{R}_{B}(M)$ the Boltzmann ROSt just defined, 
we can prove (\cite{ass}) the following
\begin{theorem}[Reversed Bound]
If we plug the Boltzmann ROSt $\mathcal{R}_{B}(M)$ 
just defined above into the 
trial function $G_N$ defined by (\ref{gtf}),
then the Boltzmann trial function $G_{N}(\mathcal{R}_{B}(M))$
provides, in the thermodynamic limit, a lower bound for the SK pressure $-\beta f$
\begin{equation*}
-\beta f\geq\lim_{N\rightarrow\infty}\liminf_{M\rightarrow\infty}
G_{N}(\mathcal{R}_{B}(M))=\lim_{N\rightarrow\infty}G_{N}(\mathcal{R}_{B})\ .
\end{equation*}
\end{theorem}
Like for CW, the idea of the proof is that 
$$
G_{N}(\mathcal{R}_{B}(M))\sim
\frac1N\mathbb{E}\ln\frac{Z_{N+M}}{Z_M} \Leftarrow M>>N\ .
$$
From the two previous theorems we get immediately the 
completion of the analogy with CW through the following 
\begin{theorem}[Extended Variational Principle] Infimizing for each $N$ 
separately the trial function $G_N(\mathcal{R})$ over the whole
ROSt space, the resulting sequence tends to the limiting pressure 
$-\beta f$ of the SK model as $N$ tends to infinity
\begin{equation*}
\label{evp-sk}
-\beta f = \lim_{N\rightarrow\infty}\inf_{\mathcal{R}}G_N\ .
\end{equation*}
\end{theorem}

Proving the reversed bound with the proper structure (the Parisi one),
is not as easy as for ferromagnets. In fact, while the magnetization is
just a number and one can fix it without too many complications like in 
(\ref{split}), the overlap is a random variable. Therefore one has to 
introduce two replicas and splitting the partition function into terms
with fixed overlap brings a lot of difficulties (\cite{t0}).

Getting back to the general approach, 
Guerra (\cite{g1}) found the following factorization property
of optimal ROSt's
\begin{theorem}\label{lisboa}
In the whole region where the parameters are uniquely
defined, the following Ces{\`a}ro  limit 
is linear in $N$ and $\bar{\alpha}$
\begin{multline}\label{limrost}
\mathbf{C}\lim_{M}\mathbb{E}\ln\Omega_M
\{\sum_{\sigma}\exp[-\beta\sum_{i=1}^N\tilde{H}_i\sigma_i
+\lambda\hat{H}]\}= \\ 
N(-\beta f 
+\frac{\beta^2}{2}(1-\langle \tilde{q}^2\rangle))+\frac{\lambda^2}{2}(1-\langle \tilde{q}^2\rangle)\ .
\end{multline}
\end{theorem}

Notice that the analogous factorization holds {\sl a fortiori} in the ferromagnetic
case, where the spin are the only variables and there is no quenched disorder
that can jeopardize the factorization.

\subsection{Generalized Random Energy Model}

The same construction we saw is reproducible in the simpler case
of the Random Energy Model and the Generalized Random Energy Model.
In this case the model is sufficiently simple to allow for a relatively
simple proof of the reversed bound. For an introduction of the model
and a detailed description of the relative ROSt approach we refer
to \cite{gs}.

\section{Spherical Models}


Let 
$$S_{N}=\{\sigma\in\mathbb{R}^{N}:
\sum_{i=1}^{N}\sigma_{i}^{2}=N\}$$
be equipped with its normalized surface measure $d\mu_{S_{N}}$.
With $J$ we will denote again a 
centered unit Gaussian random variable, and
any index or symbol appended to it will refer to independent identically 
distributed copies.
For $\sigma\in S_{N}$ the Hamiltonian of the spherical model is
the following centered Gaussian
\begin{equation*}
H=H_{N}(\sigma)=-\frac{1}{\sqrt{N}}\sum_{i,j}^{1,N} 
J_{ij}\sigma_{i}\sigma_{j}\ .
\end{equation*}
We do not consider the presence of an external field, but
all the results trivially extend to this case as well.
By $\omega$ we mean again the Bolztmann-Gibbs 
average of any observable $\mathcal{O}:S_N\to\mathbb{R}$, that 
now will be of the form
\begin{equation*}
\omega(\mathcal{O})=
Z_N^{-1}
\int_{S_{N}}\mathcal{O}(\sigma)\exp(-\beta H)d\mu_{S_{N}}\ ,\  
Z_N=\int_{S_{N}}\exp(-\beta H)d\mu_{S_{N}}\ .
\end{equation*}
Nothing changes in the expression of the internal energy when the model is
taken spherical
\begin{equation}\label{fundamental3}
\partial_\beta\frac1N \mathbb{E}\ln Z_N=\frac{\beta}{2}(1-\langle q^{2}\rangle)\ .
\end{equation}


\subsection{Generalized Bound and Extended Variational Principle}\label{general}

When the model is spherical, we still use the ROSt defined in the previous section,
except here the space $\Sigma$ in not discrete,
and the two auxiliary Hamiltonians $\tilde{H}_{.}$ and $\hat{H}$, 
defined on $\Sigma\ni \gamma$, are still Gaussian
random variables such that
\begin{equation*}
\mathbb{E}(\tilde{H}_{i}(\gamma)\tilde{H}_{j}(\gamma^{\prime}))
=N2\tilde{q}(\gamma , \gamma^{\prime})\delta_{ij}\ ,\ 
\mathbb{E}(\hat{H}(\gamma)\hat{H}(\gamma^{\prime}))=
N\tilde{q}^{2}(\gamma , \gamma^{\prime})
\end{equation*}
and $\tilde{H}=\sum_{i=1}^{N}\tilde{H}_{i}\sigma_{i}$. Then 
plug the interpolating Gaussian Hamiltonian 
$H_{t}=\sqrt{t}(H_{N}+\hat{H})+\sqrt{1-t}\tilde{H}$
into 
\begin{equation*}
R_{t}=\frac1N\mathbb{E}\ln\frac{\int_{\Sigma}\int_{\sigma}\xi_{\gamma}
\exp(-\beta H_{t})}
{\int_{\Sigma}\xi_{\gamma}\exp(-\beta \hat{H})}\ .
\end{equation*}
For most of the purposes, 
we can equivalently take either $\sigma\in\Sigma_N$ (which is the
physically natural choice, and the one we will have in mind), or
$\sigma\in\mathbb{R}^N$ equipped with
the product of independent Gaussian measures on the real line,
so we will simply write integrals over $\sigma$ without specifying
which choice we make.
We clearly have
\begin{equation}\label{interpol}
R_{0}\equiv G_{N}=\frac1N\mathbb{E}\ln\frac{\int_{\Sigma}\int_{\sigma}\xi_{\gamma}
\exp(-\beta \tilde{H})}
{\int_{\Sigma}\xi_{\gamma}\exp(-\beta \hat{H})}\ ,\ R_{1}=-\beta f_{N}\ .
\end{equation}
Now we can state the following
\begin{theorem}[Generalized Bound]\label{gbound}
The trial function $G_N=R_0$ defined in (\ref{interpol}), for any ROSt 
$\mathcal{R}$ an upper bound for the pressure, and therefore, for any $N$
\begin{equation*}
-\beta f_{N} \leq \inf_{\mathcal{R}}G_N\ .
\end{equation*}
\end{theorem}
\textbf{Proof.} Thanks to (\ref{fundamental3}), we can proceed by interpolation
like for the SK model. In fact,
\begin{equation*}
\frac{d}{dt}R_{t}=-\frac{\beta^{2}}{4}\langle(q-\tilde{q})^{2}\rangle
\leq 0\ ,
\end{equation*}
which gives the desired result because of (\ref{interpol}). $\Box$

The next definition is not exactly the expected one, 
as for simplicity we choose a handier one.
\begin{definition}
The Boltzmann ROSt $\mathcal{R}_{B}$ consists of the 
following construction
\begin{itemize}
\item the ROSt space is $\Sigma=S^\varepsilon_{M}
=\{x\in\mathbb{R}^M:|\frac1M\sum_{i=1}^Mx_i^2-1|<\varepsilon\}\ni\tau$;
\item the overlap is defined in the usual way between the spin configurations
$$
\tilde{q}(\tau,\tau^{\prime})=\frac1M\sum_{k=1}^M\tau_i\tau_i^{\prime}
$$
\item the random weights are of the Boltzmann type
\begin{equation*}
\xi_\tau=\exp(-\beta \vartheta H_{M}(\tau))\ ,\ \vartheta= \sqrt{\frac{M}{M+N}}\ ;
\end{equation*}
\end{itemize}
to be used together with the following auxiliary Hamiltonians
\begin{equation*}
\tilde{H}_{i}=-\frac{\sqrt{2}}{\sqrt{M}}\sum_{k=1}^{M} 
\tilde{J}_{ik}\tau_{k}\ ,\ 
\hat{H}=-\sqrt{\frac{N}{M}}\frac{1}{\sqrt{M}}\sum_{k,l}^{1,M} 
\hat{J}_{kl}\tau_{k}\tau_{l}\ .
\end{equation*}
\end{definition}
The choice $\varepsilon=0\Leftrightarrow \tau\in S_M=\Sigma$ 
would do as well, but it would lead
to a technically more
involved problem of equivalence of ensembles.
It will be clear that as we said we could also take the spin 
to be independent symmetric 
unit Gaussian variables, and hence take $\tau\in\mathbb{R}^M$
with the Gaussian product measure.
\begin{theorem}[Reversed Bound]
There exists an optimal ROSt that fulfills the reversed bound to 
the previous theorem, or equivalently
\begin{equation*}
-\beta f\geq 
\lim_{N\to\infty}\inf_{\mathcal{R}}G_{N}(\mathcal{R})\ .
\end{equation*}
\end{theorem}
\textbf{Proof.} We know from Talagrand's proof of the optimality of the 
Parisi ROSt $\mathcal{R}_P$ (\cite{t1}) that
$$
-\beta f=G(\mathcal{R}_P)\geq \inf_{\mathcal{R}}G_N(\mathcal{R})
$$
which is enough to prove the theorem no matter
what is the choice of the space where $\sigma$ and $\tau$
belong to, among the three mentioned possible choices. 
But we want to understand
why the statements holds from a more physical and geometrical point of view,
using the Boltzmann structure.
In a system of $M+N$ spins we can call $\tau$ 
the first $M$ and $\sigma$
the other $N$ and write
\begin{equation}\label{n+m}
H_{M+N}=-\frac{1}{\sqrt{M+N}}\left[\sum_{k,l}^{1,M} 
\hat{J}_{kl}\tau_{k}\tau_{l}+
\sqrt{2}\sum_{k\leq M,i\leq N}\tilde{J}_{ki}\tau_{k}\sigma_{i}+
\sum_{i,l}^{1,N}J_{ij}\sigma_{i}\sigma_{j}\right]\ .
\end{equation}
Thanks to the Gaussian nature of the Hamiltonians, we notice
that when $M,N\to\infty$
\begin{equation}\label{shift}
H_{M+N}\sim\vartheta(H_{M}+\tilde{H})
\end{equation}
since the third term in (\ref{n+m}) is negligible (\cite{ass}).
Moreover, again from the property of summation of independent Gaussian
variables, we have 
\begin{equation}\label{deng}
\vartheta H_{M}+\hat{H}\sim H_{M}\ .
\end{equation}
Let
$$
\Sigma^{\varepsilon}_{M+N}=\left\{ x\in\mathbb{R}^{M+N}:
\left|\frac1M\sum_{i=1}^Mx_i^2-1\right|>\frac{\varepsilon}{2} ,
\left|\frac1N\sum_{i=M+1}^{M+N}x_i^2-1\right|>\frac{\varepsilon}{2}\right\}\ .
$$
Then
\begin{multline*}
\label{ }
Z_{N+M}=\int_{\Sigma^\varepsilon_{M+N}}\exp(-\beta H_{M+N})
+\int_{S_{M+N}\setminus \Sigma^\varepsilon_{M+N}}\exp(-\beta H_{M+N})\\
\geq\int_{\Sigma^\varepsilon_{M+N}}\exp(-\beta H_{M+N})\ .
\end{multline*}
Hence we can 
proceed like in \cite{ass} and obtain
\begin{equation*}
-\beta f\geq\liminf_{N} \liminf_{M}\frac1N\mathbb{E}
\frac{Z_{M+N}}{Z_{M}}\geq G(\mathcal{R}_{B})
\geq \inf_{\mathcal{R}}G(\mathcal{R})\ .
\end{equation*}
The inequalities above rely on  
the following observations. 
The third term in (\ref{n+m}) is negligible (\cite{ass}), the second can be replaced
by $\tilde{H}$ like in (\ref{shift}), 
the first is the same as the one in $\xi_\tau$, the denominator of
$G(\mathcal{R}_{B})$ is the same as $Z_{M}$, as
guaranteed by (\ref{deng}). $\Box$ 

An intuition of why the Boltzmann ROSt works and
the useful region is where $\varepsilon$ is small is given by the following
property (\cite{will}) of the high-dimensional spheres 
$$\lim_{M,N\to\infty}
\mu_{S_{M+N}}\left(\left\{
|\frac1M\sum_{i=1}^Mx_i^2-1|>\varepsilon\right\} \bigcup
\left\{
|\frac1N\sum_{i=M+1}^{M+N}x_i^2-1|>\varepsilon\right\}\right)=0
$$

We have established, in the final $N$-limit, the reversed 
bound in Theorem \ref{gbound}
and hence also the following
\begin{theorem}[Extended Variational Principle]\label{evp-s}
Infimizing for each $N$ 
separately the trial function $G_N(\mathcal{R})$ over the whole
ROSt space, the resulting sequence tends to the limiting pressure 
$-\beta f$ of the spherical model as $N$ tends to infinity
\begin{equation*}
-\beta f = \lim_{N\rightarrow\infty}\inf_{\mathcal{R}}G_N\ .
\end{equation*}
\end{theorem}

Again, the reversed bound with the proper structure (the Parisi one),
is not as easy as for ferromagnets. In fact, the complication
is similar to that of the SK case (\cite{t1}). 



\subsection{Factorization Property}

At the beginning of this section we gave the formula of the 
generalized trial function $G_{N}$
and interpolated with the true pressure $-\beta f_{N}$. 
The spin configurations $\sigma$
were therefore constrained in $S_{N}$ and were not independent variables.
This implies that the factorization property of \cite{g1},
that we saw in the previous section for the SK model, cannot hold any longer.
But as we are interested in the thermodynamic limit, we can replace in $\tilde{H}$ 
the spins $\sigma$ by independent Gaussians (and $S_{N}$ by $\mathbb{R}^{N}$), 
and we would still get the true pressure $-\beta f$ at $t=1$ in the limit. 
We would not get $-\beta f_{N}$ at $t=1$ when $N$ is finite.
Now the cavity fields give place to independent terms and,
denoting by $dg(\cdot)$ the Gaussian measure on the real line,
the invariance of the optimal
ROSt's of Theorem \ref{lisboa} is reproduced for the 
spherical model, and we can state the following
\begin{theorem}\label{lisboa-s}
In the whole region where the parameters are uniquely
defined, the following Ces{\`a}ro  limit 
is linear in $N$ and $\lambda^{2}$
\begin{multline*}\label{limrost}
\mathbf{C}\lim_{M}\mathbb{E}\ln\Omega_M
\int_{\mathbb{R}^{N}}\prod_{i=1}^{N}dg(\sigma_{i})
\exp(-\beta\tilde{H}-\lambda\hat{H})=\\
\mathbf{C}\lim_{M}\mathbb{E}\ln\Omega_M
\prod_{i=1}^{N}\exp(\beta\tilde{H}_{i}\sigma_i)
\exp(-\lambda\hat{H})=\\
N[-\beta f 
+ \frac{\beta^{2}}{4}(1-\langle q^{2}\rangle)]+
\frac12\lambda^{2}(1-\langle q^{2}\rangle)\ .
\end{multline*}
\end{theorem}
In the last hand side above, the free energy comes 
from the fact that, like in Boltzmann ROSt,
the argument of the logarithm of $G$ is 
essentially $Z_{M+N}/Z_{M}$ exept for 
the temperature shift $\vartheta$ in (\ref{shift}), 
which gives place
to the second term. The third term is analogous to 
the second but the shift is produced
by $\lambda$. In the Boltzmann ROSt $\lambda=\beta$ 
and since $\hat{H}$ is at the
denominator of $G$ the last two terms mutually cancel 
out the only the free energy is left. The proof is therefore identical to
that of the analogous theorem for the SK model (\cite{g1}) and
similar to that of Theorem \ref{evp-s}, that is why we
only sketched it here.


\section{Dilute Spin Glasses}

Once again the spin configurations are $\sigma\in\{-1,+1\}^N$.
The Hamiltonian of the Viana-Bray (VB) model of dilute mean field spin glass is
\begin{equation*}
\label{ham}
H^{VB}_N(\sigma, \alpha; \mathcal{J})=
-\sum_{\nu=1}^{P_{\alpha N}} J_\nu \sigma_{i_\nu}\sigma_{j_\nu}\ ,\ 
\alpha\in\mathbb{R}_+
\end{equation*}
where $P_\zeta$ is a Poisson random variable of mean $\zeta$, $J_.$
are independent identically distributed copies of a random variable $J$ with
symmetric distribution, $i_.,j_.$ are  
independent identically distributed random variables with uniform
distribution over $1,\ldots,N$. The notation is the same as in the
previous sections, except $\mathbb{E}$ will be the expectation 
with respect to all the (quenched) variables, i.e. all the random variables
but the spins.
For dilute models the fundamental parameter that, by means 
of differentiation, yields the expressions analogous to 
(\ref{fundamental1})-(\ref{fundamental2})
is the degree of connectivity $\alpha$ (see \cite{lds1} for 
detailed calculations):
\begin{equation}
\label{fundamental4}
\frac{d}{d\alpha}\frac{1}{N}\mathbb{E}\ln 
\sum_{\sigma , \gamma}\xi_{\gamma}
\exp(-\beta H)
= \sum_{n>0}\frac{1}{2n}\mathbb{E}\tanh^{2n}(\beta J)
(1-\langle q^{2}_{2n}\rangle)\ ,
\end{equation}
where $q_{2n}$ is the multi-overlap defined by
\begin{equation*}
q_{2n}=\frac1N \sum_{i=1}^N\sigma_i^{(1)}\cdots\sigma_i^{(2n)}\ .
\end{equation*}
Notice that, despite the remarkable similarity with the analogous
expression in the case of non-dilute models, (\ref{fundamental4})
is not the internal energy per spin.
 
Here the key tool employed to make this calculation,
that plays the same role as the Gaussian integration by parts
for Gaussian models, is the following property of
the Poisson measure $\pi_\zeta(m)$
\begin{equation*}
\frac{d}{dt}\pi_{t\zeta}(m)=\zeta(\pi_{t\zeta}(m-1)-\pi_{t\zeta}(m))\ .
\end{equation*}
So in the various models discussed so far, the tool changes according to
the nature of the model, but the results are always pretty much alike:
quadratic in the the main physical quantity, i.e. the magnetization, the overlap,
the multi-overlap.

It should be clear now from (\ref{fundamental4}) 
that the proper structure to introduce for diluted models
is the one given in the next
 \begin{definition} 
 A {\bf Random Multi-Overlap Structure} 
$\mathcal{R}$ is a triple 
$(\Sigma, \{\tilde{q}_{2n}\}, \xi)$ where  
\begin{itemize}
  \item $\Sigma$ is a discrete space;
  \item $\xi: \Sigma\rightarrow\mathbb{R}_+$ 
  is a system of random weights; 
  \item $\tilde{q}_{2n}:\Sigma^{2n}\rightarrow[0, 1] , n\in\mathbb{N} , |\tilde{q}|\leq 1$
  is a  positive definite \emph{Multi-Overlap Kernel} 
  (equal to 1 only on the diagonal of
  $\Sigma^{2n}$),
\end{itemize}
\end{definition}
As expected, we also need to introduce the two random variables 
$\tilde{H}_{.}(\gamma, \alpha; \tilde{J})$ and $\hat{H}(\gamma, \alpha; \hat{J})$ 
such that
\begin{eqnarray}
\frac{d}{d\alpha}\mathbb{E}\ln \sum_{\gamma}\xi_{\gamma}
\exp(-\beta\tilde{H}_{.})
& = & 2\sum_{n>0}\frac{1}{2n}\mathbb{E}\tanh^{2n}(\beta J)
(1-\langle \tilde{q}_{2n}\rangle)\label{eta} \\
\frac{d}{d\alpha}\frac 1N\mathbb{E}\ln \sum_{\gamma}\xi_{\gamma}
\exp(-\beta \hat{H})
&=& \sum_{n>0}\frac{1}{2n}\mathbb{E}\tanh^{2n}(\beta J)
(1-\langle\tilde{q}^{2}_{2n}\rangle)\label{kappa}
\end{eqnarray}
to be used in the usual trial function
\begin{equation}\label{gtfdilute}
G_{N}(\mathcal{R})=\frac 1N\mathbb{E}\ln
\frac{\sum_{\sigma, \tau}\xi_{\tau}
\exp(-\beta\sum_{i=1}^{N}\tilde{H}_{i}\sigma_{i})}{\sum_{\tau}\xi_{\tau}
\exp(-\beta \hat{H})}
\end{equation}
where $\tilde{H}_{i}$ are independent copies of $\tilde{H}_{.}$.
The identity (\ref{fundamental4}) also suggests to interpolate on the
connectivity in order to prove (\cite{lds1}) 
the following analog of Theorems \ref{b1} and \ref{b2}.
\begin{theorem}[Generalized Bound]
\label{b3}
For any 
value of $N\in\mathbb{N}$, the trial function $G_N$ defined 
by (\ref{gtfdilute}) is an upper bound
for the pressure $-\beta f_N$ of the VB model
\begin{equation*}
\label{ }
-\beta f\leq \lim_{N\rightarrow\infty}
\inf_{\mathcal{R}} G_N(\mathcal{R})\ .
\end{equation*}
Again, from (\ref{fundamental4}) we learn that the generalized bound
is in the end equivalent to the extension to all even multi-overlap of (\ref{main2})
\begin{equation}
\label{main3}
q^2_{2n} \leq 2q_{2n}\tilde{q}_{2n} - \tilde{q}^2_{2n}\ \ \forall\ n\in\mathbb{N}_0\ .
\end{equation}

The Boltzmann RaMOSt (\cite{lds1}) is defined in total analogy
with the one in the previous section
$$
\Sigma=\{-1,1\}^M\ni\tau\ ,\ \xi_\tau=\exp(-\beta H_M)\ ,\ 
\tilde{q}_{1\cdots 2n}=\frac1M\sum_{k=1}^M\tau^{(1)}_k\tau^{(2n)}_k 
$$
with
\begin{eqnarray}
\label{hat}
&&\tilde{H}=
-\sum_{\nu=1}^{P_{2\alpha N}} 
\tilde{J}_\nu \tau_{k_\nu}\sigma_{i_\nu}\equiv 
\sum_{j=1}^{N}H_{i}\sigma_{i}\ , 
H_{i}=\sum_{\nu=1}^{P_{2\alpha}}J_{\nu}^{i}\tau_{k_{\nu}^{i}}\\
\label{tilde}
&&
\hat{H}=
-\sum_{\nu=1}^{P_{\alpha N}} 
\hat{J}_\nu \tau_{k_\nu}\tau_{l_\nu}\ , 
\end{eqnarray}
and again it yields 
\begin{theorem}[Reversed Bound]
If we plug the Boltzmann ROSt $\mathcal{R}_{B}(M)$ 
just defined above into the 
trial function $G_N$ defined by (\ref{gtfdilute}),
then the Boltzmann trial function $G_{N}(\mathcal{R}_{B}(M))$
provides, in the thermodynamic limit, a lower 
bound for the VB pressure $-\beta f$
\begin{equation*}
-\beta f\geq\lim_{N\rightarrow\infty}\liminf_{M\rightarrow\infty}
G_{N}(\mathcal{R}_{B}(M))=\lim_{N\rightarrow\infty}G_{N}(\mathcal{R}_{B})\ .
\end{equation*}
\end{theorem}
The obvious consequence is again
\end{theorem}
\begin{theorem}[Extended Variational Principle]
Infimizing for each $N$ 
separately the trial function $G_N(\mathcal{R})$ over the whole
ROSt space, the resulting sequence tends to the limiting pressure 
$-\beta f$ of the SK model as $N$ tends to infinity
\begin{equation*}
\label{evp-d}
-\beta f=\lim_{N\rightarrow\infty}
\inf_{\mathcal{R}}G_{N}(\mathcal{R})\ .
\end{equation*}
The physics of dilute models is still quite obscure, and the technical
difficulties prohibitive (as of now). The rigorous calculation of the free
energy is still missing. Still, as we showed, the structure of dilute models
is similar to the one of the others. 
See \cite{franz, pt, lds2} for recent progress on the subject.

The analogy with SK is completed by the next (\cite{lds1}) result
\end{theorem}
\begin{theorem}[Factorization of optimal RaMOSt's]
\label{lisboa-d}
In the whole region where the parameters are uniquely
defined, the following Ces{\`a}ro  limit 
is linear in $N$ and $\bar{\alpha}$
\begin{equation*}\label{limrost}
\mathbf{C}\lim_{M}\mathbb{E}\ln\Omega_M
\{\sum_{\sigma}\exp[-\beta(\tilde{H}(\alpha)+\hat{H}(\bar{\alpha}/N))]\}
=N(-\beta f +\alpha A)+\bar{\alpha}A\ ,
\end{equation*}
where
\begin{equation*}
\label{ }
A=\sum_{n=1}^{\infty}\frac{1}{2n}
\mathbb{E}\tanh^{2n}(\beta J)(1-\langle q_{2n}^{2}\rangle)\ ,\ 
\tilde{H}=\sum_{i=1}^{N}\tilde{H}_{i}\sigma_{i}\ .
\end{equation*}
\end{theorem}

\subsection{Optimization Problems}

We conclude by briefly describing how all we just saw extends to 
optimization problems as well.

The Hamiltonian of the random K-SAT is a model of dilute
spin glasses, that using the usual notations reads
\begin{equation*}
H=-\sum_{\nu=1}^{P_{\alpha N}} \frac12(1+J^{1}_\nu\sigma_{i^{1}_\nu})
\cdots\frac12(1+J^{K}_\nu\sigma_{i^{K}_\nu})\ ,
\end{equation*}
where $\{i^{\mu}_\nu\}$ are independent identically 
distributed random variables, 
uniformly distributed over points $\{1,\ldots, N\}$,
$\{J^{\mu}_\nu\}$ are independent identically 
distributed copies of a symmetric random variable $J=\pm 1$, 

The fundamental equation we get this time (\cite{lds3}) is
\begin{equation}\label{basic}
\frac{d}{d\alpha}\frac 1N\mathbb{E}\ln \sum_{\gamma}\xi_{\gamma}
\exp(-\beta H)
 =  \sum_{n>0}\frac{(-1)^{n+1}}{n}\left(\frac{e^{-\beta}-1}{2^{K}}\right)^{n}
\langle (1+Q_{n}(q))^{K}\rangle
\end{equation}
where 
\begin{equation*}
Q_{2n}(q)=\sum_{l=1}^{n}
  \sum^{1,2n}_{r_{1}<\cdots< r_{2l}}q_{r_{1}\cdots r_{2l}}\ ,\ 
  Q_{2n+1}=0\ ,\ n=1, 2, \ldots
\end{equation*}
which explain why we can still the RaMOSt, provided we consider
$\tilde{q}_{r_{1}\cdots r_{2l}}:\Sigma_{r_{1}}
\times\cdots\times\Sigma_{r_{l}}\rightarrow[0, 1]$
instead of the mere even multi-overlap as for the Viana-Bray model.

The natural modification of the usual quantities is (\cite{lds3}) thus
\begin{eqnarray*}
&&\hspace{-0.7cm}\frac{d}{d\alpha}\frac 1N\mathbb{E}\ln \sum_{\gamma}\xi_{\gamma}
\exp(-\beta\tilde{H}_{.}) 
= K\sum_{n>0}\frac{(-1)^{n+1}}{n}\left(\frac{e^{-\beta}-1}{2^{K-1}}\right)^{n}\!\!
\langle (1+Q_{n}(\tilde{q}))^{K-1}\rangle \\
&&\hspace{-0.7cm}\frac{d}{d\alpha}\frac 1N\mathbb{E}\ln \sum_{\gamma}\xi_{\gamma}
\exp(-\beta \hat{H})
= (K\!-\!1)\sum_{n>0}\frac{(-1)^{n+1}}{n}\left(\frac{e^{-\beta}-1}{2^{K}}\right)^{n}\!\!
\langle (1+Q_{n}(\tilde{q}))^{K}\rangle
\end{eqnarray*}
and
\begin{equation*}
G_{N}(\mathcal{R}, \tilde{H}, \hat{H})=\frac 1N\mathbb{E}\ln
\frac{\sum_{\sigma, \tau}\xi_{\tau}
\exp(-\beta\sum_{i=1}^{N}(\tilde{H}_{i})
\frac12(1+J_{i}\sigma_{i}))}{\sum_{\tau}\xi_{\tau}
\exp(-\beta \hat{H})}
\end{equation*}
where $\tilde{H}_{i}$ are independent copies of $\tilde{H}_{.}$, and finally
\begin{equation*}
\tilde{H}=\sum_{i=1}^{N}\tilde{H}_{i}\frac12(1+J_{i}\sigma_{i})\ .
\end{equation*}
The same interpolation on the connectivities allows one to prove the generalized
bound, which is equivalent to the non-negativity of the function 
$x^{K}-Kxy^{K-1}+(K-1)y^{K}$ 
of $x$ and $y$.
The obvious construction of the Boltzmann RaMOSt (\cite{lds3}) is 
$\Sigma=\{-1,1\}^{M}$, using $\tau$ instead of $\gamma$,  
$\xi_{\tau}=\exp(-\beta H_{M}(\tau))$ and
\begin{eqnarray*}
\tilde{H}_{\tau} &=&
-\sum_{\nu=1}^{P_{K\alpha N}}
\frac12(1+\tilde{J}^{1}_\nu\tau_{j^{1}_\nu})
\cdots\frac12(1+\tilde{J}^{K-1}_\nu\tau_{j^{K-1}_\nu})
\frac12(1+J^{K}_\nu\sigma_{i_\nu})\\
\hat{H}_{\tau} &=&
-\sum_{\nu=1}^{P_{(K-1)\alpha N}}
\frac12(1+\hat{J}^{1}_\nu\tau_{j^{1}_\nu})
\cdots\frac12(1+\hat{J}^{K}_\nu\tau_{j^{K}_\nu})
\end{eqnarray*}
where the independent random variables $j_{.}^{.}$ are all
uniformly distributed over $1,\ldots,M$
and $\tilde{J}_{.}^{.}, \hat{J}_{.}^{.}$ are independent copies of $J$.
Of course the Boltzmann RaMOSt fulfills the reversed bound and
leads therefore to 
the extended variational principle (\cite{lds3}).
Lastly the same factorization property as usual is verified here as well (\cite{lds3}).


\section{Conclusion}

If we start from the approach of \cite{ass} for the SK model, we 
saw that it keeps its validity if we remove the disorder (passing to the CW model),
if we dilute the model (passing to the VB), if we make the model spherical.
The presence of an external field can easily taken into account in all cases.
Unfortunately, the approach does not directly extend to the dilute case 
if the model is spherical or ferromagnetic (some interesting results
can be obtained in this latter case, and we plan on reporting on this soon
\cite{lds6}). Moreover, it is not clear what
to do when interactions involve an odd number of spins. But the main
open problem is probably to understand which is the right extension to
finite dimensional models.


\section*{Acknowledgments}

The author is grateful to Lorenzo Bertini for useful discussions and warmly thanks Francesco Guerra for so many precious conversations. Many thanks are also due
to an anonymous referee for totally appreciated suggestions.


\end{document}